# Revealing the spatial extent of patent citations in the UK: How far does knowledge really spillover?

Philip Wilkinson[1] and Elsa Arcaute[1]

[1]Centre for Advanced Spatial Analysis, University College London, 90 Tottenham Court Road, London, W1T 4TJ, UK

**Abstract**

Access to external knowledge sources through localized knowledge spillovers is an important determinant of the innovative capabilities of firms. However, the geographical extent of knowledge spillovers is not well understood. In this article we use patent citations in the UK as a proxy of knowledge flows and analyze the spatial extent of knowledge spillovers relative to the distribution of existing knowledge creation. We find that local, regional and country specific institutional factors play an important role in influencing the probability of knowledge spillovers and that most knowledge spillovers are exhausted within an extended commuting boundary. It is also shown that these effects have increased over time and that the spatial extent of knowledge spillovers varies by industry.

## 1 Introduction

Innovation depends on the combination of existing ideas, concepts, and theories to create something new which can be utilised for societal or commercial benefit (Carlino & Kerr, 2015). To this end it is increasingly recognised that the innovative capacity of a firm depends not only on its ability to combine internal stocks of knowledge and resources but also its ability to acquire and implement external sources of knowledge (Paci, et al., 2014; Crescenzi, et al., 2016). Utilising these external sources reduces the internal costs of innovation and allows firms to develop novel combinations which they may not have developed on their own (Rammer, et al., 2020). In this sense, geographical proximity has been identified as a factor that can influence the access to these external resources through knowledge spillovers, affecting the innovative performance of firms. However, while this has been acknowledged, little is known about the exact geographic scope of these spillovers (Cappelli, et al., 2014).

The idea of knowledge spillovers was originally conceived by Alfred Marshall as 'The mysteries of the trade become no mysteries; but are as if it were in the air' (Marshall, 1890, p. 251), which is similarly reflected in subsequent theories from a variety of academic disciplines (McCann, 2013). The importance of this, and its dynamic understanding, can be readily seen in recent developments of the literature that focus on the agglomeration of firms, such as Endogenous Growth Theory and Porter's Cluster Concept (Fritsch & Franke, 2004). Different factors have been suggested to influence these knowledge spillovers such as physical distance between firms, institutional difference and the degree of knowledge crossover, however all three have not been considered in combination. In this piece we examine these factors concurrently, considering the physical distance between inventors/applicants, the institutional constraints through local, regional and country level boundaries, and knowledge similarity related to the technological domain.

Different proxies have been used in the existing literature to quantify these knowledge spillovers, such as wage rates, worker mobility, industry concentration, the flow of goods and patents. Given the factors that we want to examine, the influence of one innovation on another relative to their distance, institutional membership and industrial similarity, patents can provide highly granular spatial information (Belenzon & Schankerman, 2013; Buzard, et al., 2020). This is in addition to reflecting the communication efforts between firms and their relationship with innovative outcomes (Jaffe, et al., 2000; Duguet & MacGarvie, 2005), suggesting a strong relationship to knowledge spillovers. Their use however is not without caveats. The most important one refers to the fact that patents are registered at the headquarters of the company/institution rather than at the location of the innovation (Moreno, et al., 2005). We address this issue, by considering the address of the applicant/inventor instead of the firm's address. This provides a better estimate for exploring the effect of distance on the probability of citations.

Although patents have been used in the past to explore knowledge spillovers, most work so far has focused on the United States (Howells, 2002). Building on previous literature that has begun to explore the European example (Duguet & MacGarvie, 2005; Moreno, et al., 2005; Helmers, 2019) this work focuses on the UK system through patent citations data from the OECD REGPAT database. We investigate the effects of technological overlap, administrative boundaries and geographical distance simultaneously through a binomial t-test and Probit regression methodology. The results show that local (NUTS3), regional (NUTS1) and country boundaries are positively associated with knowledge spillovers, and that most knowledge spillovers are exhausted within a 100-150 mile range, suggesting that face-to-face interaction is crucial for knowledge transfer. Furthermore, the effects of distance and administrative boundaries on knowledge spillovers have become stronger over the period of 1977-2019, meaning that in general, there is no evidence that technology has reduced the dependence on face-to-face interactions but has, in fact, increased it. Finally, geography affects knowledge spillovers differently across industries, with the Pharmaceutical and IT industries exhibiting strong distance-based effects, while Biotechnology and Medical industries appear to have weaker or even non-significant distance-based effects in the UK. These findings have important implications for industrial and innovation policy aiming to generate economic growth through the transfer of knowledge between sectors and firms.

The paper is organised as follows. In section 2 we provide an overview of agglomeration economies and knowledge spillovers in order to motivate our hypotheses. Details and exploration of the data can be found in section 3. The methodology is presented in section 4. The analyses and results of the effects of distance and administrative boundaries in relation to the hypotheses developed in section 2 are presented in section 5. Finally, section 6 contains the conclusions and reflections on the work.

## 2 Agglomeration economies and knowledge spillovers

The concept of agglomeration economies has enjoyed a revival in the literature since the beginning of the 1980s, associated with an increased focus on the role of dynamic influences such as social, cultural and political factors that play a role in cluster formation, development and subsequent decline (Malmberg, 1996). The re-emergence and change in focus occurred due to a shift towards institutional and evolutionary issues within economics, the recognition of the significant fall in transport costs during the 20th Century, and the growing importance of the knowledge-based economy in modern economic growth (Malmberg & Maskell, 2002; Mackinnon & Cumbers, 2011). This led to the development of new theories and concepts including the development of New Economic and Endogenous Growth Theories, the advancement and development of evolutionary and institutional approaches within economics (McCann & Van Oort, 2019), and the creation of multiple different cluster concepts from a variety of disciplines (Gordon & McCann, 2000; Martin & Sunley, 2003). However, a consistent factor found in these theories and concepts is that of knowledge generation, accumulation and spillovers which are suggested to be one of the key forces driving cluster benefits, including innovation (McCann & Van Oort, 2019). This has developed in tandem with innovation literature that has placed an increased emphasis on access to external knowledge sources to help drive internal innovation (Herstad, et al., 2014).

It is recognised that measuring knowledge spillovers is difficult, as emphasised by Krugman (1991, p.53) that 'knowledge flows … leave no paper trail by which they may be measured and tracked'. Thus, any method attempting to do so must measure them indirectly (Thompson, 2006) leading to a variety of different methodologies applied. This has included the use of wages (Glaeser, 1999; Glaeser & Mare, 2001), tracking movements between firms (Breschi & Lissoni, 2000; Miguelez & Moreno, 2015), and the trade of goods between regions (Kogler, 2015). However, these measures suffer from the issue such that they often track pecuniary externalities, rather than pure knowledge spillovers (Henderson, 2007), and that much of their data is available at a highly aggregated level (Feldman & Kogler, 2010). Patents on the other hand, contain a high degree of granularity and the potential for knowledge spillovers to be tracked through citations (Feldman & Kogler, 2010). Furthermore, the application for a patent is often costly and time consuming such that their value must exceed the costs of the application making them a valid proxy for innovation (Moreno, et al., 2005). Consequently, the use of patent and patent citations for this purpose has followed their digital publication in the late 1990s (Carlino & Kerr, 2015).

In terms of knowledge spillovers, patents can be used as a proxy for the paper trail of knowledge as they contain information on, including the location of, the firm and the inventor, along with previous patents that they must be related to (Thompson & Fox-Kean, 2005). Jaffe et al. (2000) in the case of the US, and Duguet and MacGarvie (2005) in the case of Europe, find evidence of communication and knowledge flows between citing and cited patent inventors through informal communication or technologies suggesting that citations do indicate evidence of knowledge flows. This is even so in the case of European Patent Office where citations can be added by examiners (Webb, et al., 2005), thus it can still be assumed that inventor teams would have likely been aware of these original patents (Maurseth & Verspagen, 2002). Furthermore, in the case of the US, evidence suggests that potential knowledge flows embedded in citations are geographically localised (Belenzon & Schankerman, 2013; Singh & Marx, 2013; Buzard, et al., 2020), which has also been supported for European patents (Maurseth & Verspagen, 2002; Fischer, et al., 2009).

However, issues of using this measure must also be recognised. This includes the fact that not all industries patent which makes it difficult to generalise the results (O'Clery, et al., 2019), not all inventions that can be patent, are patented, due to the requirement of full disclosure whereby trade secrets may be better protected internally (Agrawal, et al., 2014), patents represent codified knowledge rather than tacit (Howells, 2002), and that other sources of knowledge spillovers or transfers are not captured by patent citations (Dechezleprêtre, et al., 2013). Despite this, they are accepted as one of the best mechanisms for measuring knowledge spillovers and thus widely used in academic research for now (Buzard, et al., 2020).

Most studies on the subject of knowledge spillovers explored through the use of patents examine the case of the United Stated (Howells, 2002; Belenzon & Schankerman, 2013; Singh & Marx, 2013; Buzard, et al., 2020), with only a few exploring the example of Europe (Maurseth & Verspagen, 2002; Fischer, et al., 2009; Helmers, 2019). This paper seeks to extend this corpus of literature by examining the case of the UK in terms of the effects that geography may have on knowledge spillovers. The purpose of this is to be able to utilise a new case study and derive implications for existing innovation policy while opening up the debate on existing theory and results.

In doing so, several different hypotheses are explored in terms of the influence of geography on knowledge spillovers. First, in terms of the role of administrative boundaries, such as the metropolitan statistical areas, regions and country level in the US, early research using patents find statistical significant results of geographically bounded spillovers, subsequently supported by further research (Jaffe, et al., 1993; Belenzon & Schankerman, 2013; Singh & Marx, 2013). However, beyond geography, this effect is also suggested to be due to the way institutions function, where informal codes of interaction, technical language usage and worker mobility can support the process of knowledge spillovers within administrative boundaries (Malmberg, 1996; Agrawal, et al., 2014; Speldekamp, et al., 2020). Thus, it is reasonable to assume that such findings are also seen in the UK, despite differences in the scale of the local, regional and country level.

> *Hypothesis 1:* Administrative boundaries will have a positive and significant effect on the probability of citation at the local (NUTS3), regional (NUTS1) and country level in the UK.

The second question builds on this by seeking to explore explicitly the distance over which knowledge spillovers operate in the UK. This is important as administrative boundaries act only as a proxy for distance, thus exploring pure distance and administrative boundaries simultaneously could sperate the institutional from the distance effects. This is a key question that is still being examined in the US, with research finding evidence of knowledge spillovers being exhausted within an extended commuting boundary of 50-150 miles (Belenzon & Schankerman, 2013; Singh & Marx, 2013), within MSA distance of 20 miles (Buzard, et al., 2020), or even within a few hundred metres (Arzaghi & Henderson, 2008). Thus, examining this in the UK will add to the debate as to the extent of spatial spillovers. Here, we examine this by considering the NUTS3 boundaries of the UK, whereby knowledge spillovers are expected to be exhausted within a scale of 0-100 miles.

> *Hypothesis 2:* Most of the knowledge spillovers will be exhausted within a range of 0-100 miles.

It is also important to examine how the effects of distance has changed over time. This relates to the questions of whether new communication technologies have actually led to the 'death of distance' given their ability to facilitate regular long distance communication (Sonn & Stoprer, 2008). This is especially important for knowledge sharing for the purposes of innovation. While there are still costs for the transportation of physical goods in terms of time and money, ideas are thought to be instantly transferrable through mediums such as email, telephone or the internet. Thus, if technology can replace face-to-face interaction for knowledge sharing, then we would expect to see the effects of both distance and administrative boundaries decreasing (Howells, 2002; Keller, 2002; Thompson, 2006). This result would be especially important given the rise of work from home due to the current COVID-19 pandemic and predicted continuance of this trend (Hern, 2020; Eisenberg, 2020; Regan & Harby, 2020), limiting face-to-face contact.

> *Hypothesis 3:* The strength and importance of administrative boundaries and geographic distance in knowledge spillovers will decrease over time.

Finally, it also important to explore the different effects that distance and administrative boundaries may have across different industries (McCann & Van Oort, 2019). There is evidence that agglomeration forces, knowledge spillovers and innovative outcomes across industries differ. For example, in the UK Duranton and Overman (2005) examined industry location and found that while for some industries such as publishing or chemicals, localisation was important within scales of 0-50km others such as machinery for textiles were also important at scales of 80-140km. Using patents, Belenzon and Schankerman (2013) in the US found that the effects of distance were less pronounced in Biotechnology, IT and Telecommunications compared to Chemical and Pharmaceutical industries, while Adams and Jaffe (1996) found that spatial proximity was significant in the Pharmaceutical industry. It is suggested that this is because of different technologies used for communication across industries and also different standards of codification or accepted knowledge sharing practices, leading to separate emphasis on codified or tacit knowledge sharing within different domains. Of course this could also be related to the different propensity to patent across industries (O'Clery, et al., 2019), however the fact that different results are found for the same industries at different points using different methodologies, suggests there is still some ambiguity to be explored.

> *Hypothesis 4*: Knowledge spillovers will be affected differently across industries by administrative and distance boundaries.

# 3 Patent data exploration

Patent data comes from the OECD REGPAT database, containing information on patents applied for under the European Patent Office (EPO) and the Patent Co-operation Treaty (PCT) (OECD, January 2020). This includes the application number, original application year and their International Patent Classification which shows the patents industry domain. This also includes the individual's and firm's address, country code and their share of the application. From this, patents whose entire investor team was located in the UK were extracted to serve as the original patents. This is to prevent uncertainty with respect to assigning the correct country for the innovation when the patent is the result of an international team (Belenzon & Schankerman, 2013). Furthermore, when identifying the location of the patent, the inventor's address was used, rather than the firm's, because firms often register patents at their headquarters, affecting the calculation of distances (Moreno, et al., 2005). While this may limit the exploration of clusters of firms, it provides a more reasonable estimate of the geography over which knowledge spillovers occur (Buzard, et al., 2020).

In order to test for evidence of the effects of geography, the distribution of existing knowledge production had to be controlled for (Duranton & Overman, 2005). For this, a control sample of patents were created, matching the four digit IPC code and priority year of the citing patent, but that did not cite the original patent (Thompson & Fox-Kean, 2005). From this, a random control patent was extracted for each cited/citing pair, to compare the actual distribution of knowledge spillovers against the potential distribution of knowledge production (Carlino & Kerr, 2015). This follows the well-established matching method developed by Jaffe et al. (1993), and allows for the comparison of the actual citations with respect to potential citations in terms of geography (Kogler, 2015). In addition, we also acknowledge the critique of Thompson and Fox-Kean (2005) for Jaffe et al. (1993) of not matching at enough of the IPC digit codes by also including a control for a six-digit IPC code match within the regression formulation.

Distance between cited and citing/control pairs was calculated using the Haversine formula based on the latitude and longitude associated with the local regions assigned. Mean distance between teams was taken as the geographical distance over which the spillover may have occurred (Belenzon & Schankerman, 2013), because we do not know how knowledge was transmitted. By doing so, we are reducing the bias of assuming that knowledge travels through the connection related to the shortest distance. This can nevertheless, bias the results, where no geographical effect is found. Despite this possible bias, we do find evidence of localised knowledge spillovers. Figure 2 and Table 1 below highlight the difference between the treatment and control dataset created.

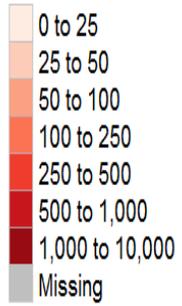
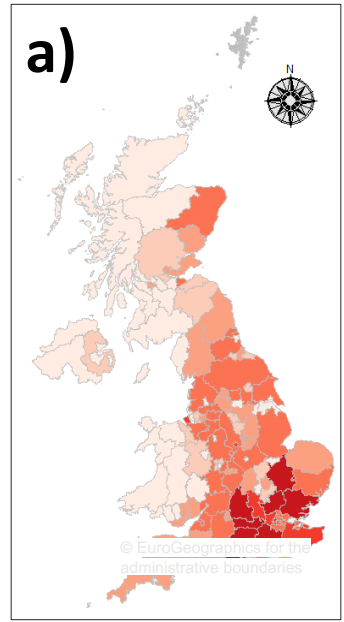
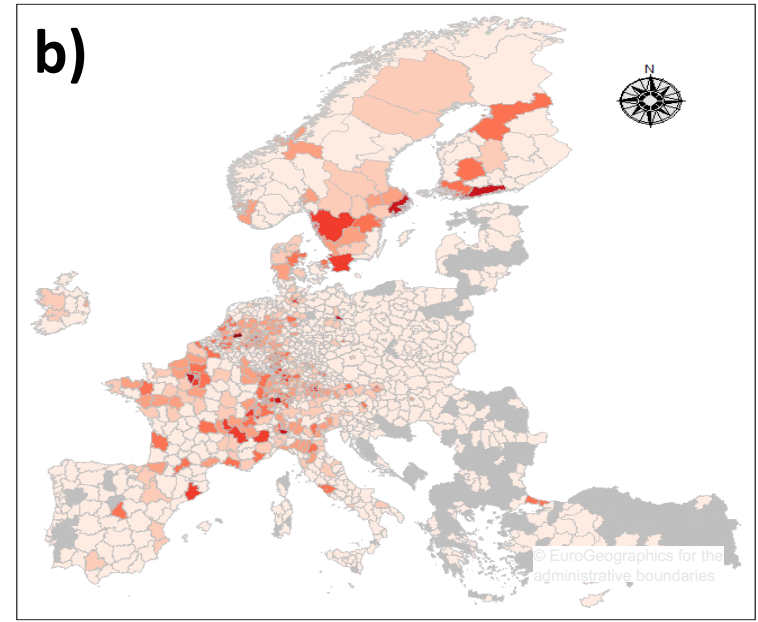
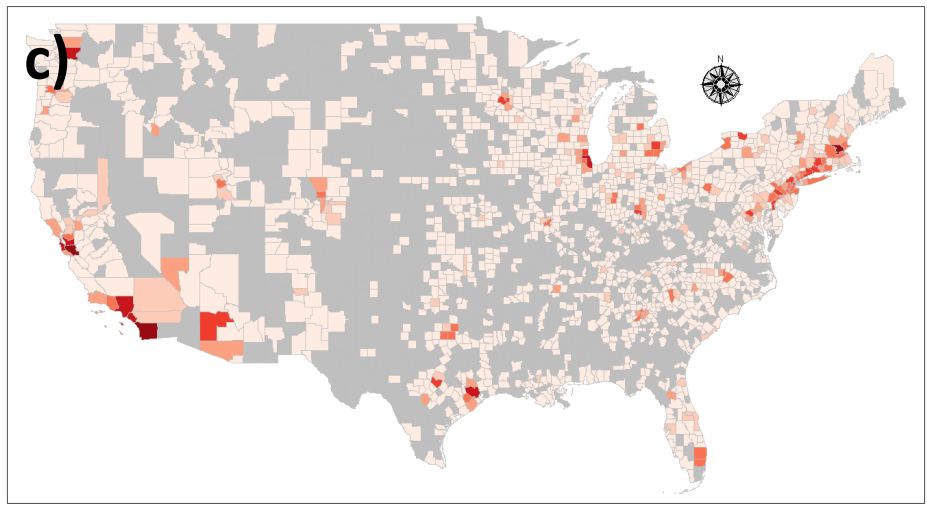
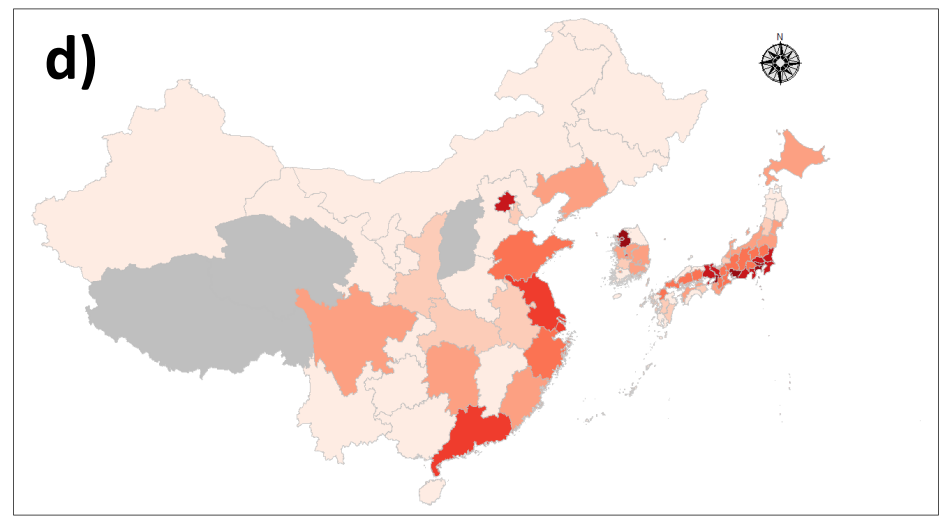

*Figure 1* - The regional distribution of the sum of patent inventor share by individuals from 1977 to 2019 from patents that cite UK patents. The regions that produce the most citations are shown here with a) UK, b) Europe, c) United States, d) China, South Korea and Japan.

*Figure 1* - Percentage difference between actual and control citations at distance boundaries (%)

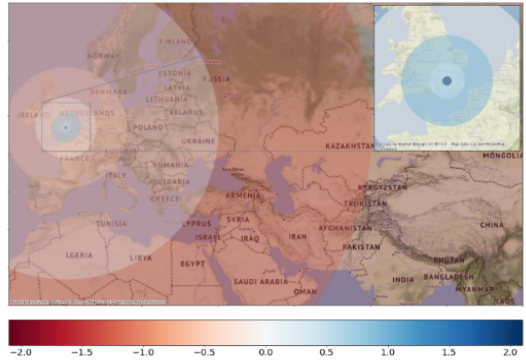

*Table 1*: Boundary, distance and IPC matching percentages of citations and control

|  | **Treatment** | | **Control** | | |
| --- | --- | --- | --- | --- | --- |
|  | Percentage | Count | Percentage | Count | Percentage difference |
| **Administrative boundaries** | | | | | |
| Local match | 2.84% | 4140 | 0.37% | 542 | 2.47% |
| Regional match | 4.54% | 6607 | 1.20% | 1741 | 3.34% |
| Country match | 9.99% | 14556 | 5.63% | 8205 | 4.36% |
| **Distance boundaries (miles)** | | | | | |
| 0 to 25 | 2.44% | 3559 | 0.37% | 541 | 2.07% |
| 25 to 50 | 1.16% | 1690 | 0.60% | 875 | 0.56% |
| 50 to 100 | 1.78% | 2593 | 1.26% | 1838 | 0.52% |
| 100 to 250 | 5.09% | 7416 | 4.15% | 6048 | 0.94% |
| 250 to 500 | 13.70% | 19950 | 13.76% | 20046 | -0.07% |
| 500 to 1000 | 17.83% | 25967 | 18.43% | 26839 | -0.60% |
| 1000 to 2500 | 4.49% | 6539 | 4.24% | 6182 | 0.25% |
| 2500 to 5000 | 25.59% | 37262 | 26.86% | 39122 | -1.28% |
| > 5000 | 27.92% | 40657 | 30.31% | 44142 | -2.39% |
| **6-digit IPC match** | 22.68% | 33035 | 14.25% | 20749 | 8.44% |

# 4 Binomial t-test and Probit Regression

To test for localisation at the local (NUTS3), regional (NUTS1) and country level, a binomial t-test is used (Jaffe, et al., 1993). We explore whether the regions assigned to the original and citing/control patents are the same using the formula below:

$$t = \frac{P_c - P_o}{\sqrt{(P_c(1-P_c) + P_o(1-P_o))/n}} \quad (1)$$

The test is H$_0$: P$_c$ = P$_o$ against H$_a$: P$_c$ > P$_o$ where P$_c$ is the proportion of actual citations that come from the same geographical unit that the original patent comes from, P$_o$ being the proportion of control citations that come from within the geographical boundary, and n representing the number of cited-citing pairs (Jaffe & Trajtenberg, 1996). The t-value is used to calculate the significance of this result, providing evidence of localisation at the given scales under analysis.

Although the test can provide some insights, this is insufficient given the following caveats: 1) adjacent administrative units are treated the same as units that are at the opposite end of the country and therefore may underestimate the effect of knowledge spillovers (Murata, et al., 2014); 2) the test can only be applied for one boundary condition at a time (Singh & Marx, 2013); 3) administrative boundaries do not always reflect economic boundaries and therefore distance must also be considered (Duranton & Overman, 2005). We Therefore utilise a Probit regression in addition to this.

The dichotomous dependent variable of whether the control/citing patent did cite the original patent is modelled as follows:

$$P(Y_{ij} = 1|D_{ij}, D_{wb}, T_{ij}) = \Phi(\beta_0 + \beta_1 D_{ij} + \beta_2 D_{wb} + \beta_3 T_{ij} + \epsilon_{ij}) \quad (2)$$

where $Y_{ij}$ is a dummy variable equal to 1 if patent *i* is cited by patent *j* and 0 otherwise, $D_{ij}$ are a set of dummies for distance boundaries between *i* and *j*, $D_{wb}$ are a set of border dummy's to account for the same administrative border between patents *i* and *j*, and $T_{ij}$ is a dummy variable indicating if there is a match at the 6-digit IPC level between *i* and *j*. Furthermore, $\epsilon$ is given as the error term and $\Phi$ represents the Probit link function (Talibova, et al., 2018). Unlike the binomial t-test this model controls for distance and boundary effects simultaneously, along with a greater precision of technology matching. The latter allows us to consider that knowledge is more likely to spillover within technology subclasses than between them, (see critique by Thompson and Fox-Kean (2005)) thus requiring a greater degree of specification of patent class. From this, the margin coefficients are presented which represent the percentage point change in the probability that the dependent variable would equal one if, *ceteris paribus*, the independent dummy variable changes from zero to one (Leeper, 2018).

# 5 Results
## 5.1 Binomial t-test results

The purpose of this test is to test for evidence of localisation at the local (NUTS3), regional (NUTS1) and country level individually. This does so by controlling for the existing distribution of knowledge through the control group such that if the actual citation matching probability within an administrative unit exceeds that of the control group, then this suggests that knowledge spillovers are localised. To this extent, evidence found in table 3 below suggests that knowledge spillovers are localised within each administrative boundary as the actual citation matches at each level are statistically significantly higher than the matching percentage in the control dataset.

*Table 2: Binomial t-test results for administrative boundaries*

| Administrative boundary | Count | Actual citation match (%) | Control citation match | Difference in percentage values | T-statistic |
|---|---|---|---|---|---|
| Local | 291044 | 2.84% | 0.37% | 2.47% | (53.21)*** |
| Regional | 291044 | 4.53% | 1.19% | 3.33% | (54.21)*** |
| Country | 291044 | 9.99% | 5.63% | 4.35% | (43.89)*** |

*This table reports the results of a binomial t-test using equation 1, comparing the matching of the treatment and control patents within each administrative boundary to the cited patent. Significance level is represented by \*, \* = 10%, \*\* = 5%, \*\*\* = 1%.*

This therefore suggests evidence in support of hypothesis 1, that similar to early work with patent analysis, knowledge spillovers are constrained within local administrative boundaries (Jaffe, et al., 1993; Belenzon & Schankerman, 2013; Singh & Marx, 2013). If taken as a proxy of distance, this would provide evidence that knowledge spillovers are constrained by distance. However, due to the different sizes of administrative units we are unable to tell definitively over which scale do knowledge spillovers act and within administrative units institutional factors may also limit knowledge spillovers between firms. Thus, we need to be able to control for actual distance, as well as administrative units, to be able to separate out these effects.

## 5.2 Baseline regression estimation results

The Probit regression methodology allows us to do so, and the results can be found in table 3 below. Firstly, column 1 examines the effects of boundaries which show evidence of knowledge spillovers constraints, but with the strength of this effect decreasing with the increasing size of the boundary, supporting the results of the binomial t-test and hence hypothesis 1. This is because it shows that a citation is much more likely to come from a patent whose team is located within the same local administrative unit than the distribution of knowledge would suggest, controlling for both regional and country level effects. The initial effects is associated with an increase of 26.7 percentage point increase in the probability of citation, which accounts for a 50% increase in the baseline probability of citations (50%). Region and Country level effects are then associated with a 12.1 and 5.4 pp increase in the probability of citation.

Column 2 and 3 therefore seek to examine the effects of distance alone. Here, column 2 utilising log distance shows has a negative and significant coefficient, suggesting that as distance increase the probability of citation decreases. However, this does not show the distance over which knowledge spillovers operate. Therefore, with the expectation that knowledge spillovers decay exponentially, the second column shows the effects of distance through intervals. Here, 0-25 miles is taken as the reference condition and so each distance interval is compared to this. As a result, the 25-50 mile boundary shows a reduction in the probability of citation by 25.7 percentage points, a 50% decrease against the 50% baseline probability of citation. Moving then to the 50-100 mile boundary leads to a further reduction of 6.9 percentage points, and the 100-250 mile boundary a further drop of 3.3. percentage points. By the 250-500 mile boundary, the total loss is 40.7 percentage points, after which distance has no more depreciable effect on citation, associated with an overall reduction of 80% against the baseline probability of citation. This therefore supports the argument that knowledge spillovers are mostly exhausted within the 0-100 mile range, but also that knowledge spillovers may extend beyond this range to some extent. However, this does not control for both distance and administrative boundaries simultaneously and so extent of knowledge spillovers may be overstated.

Column 4 is used to show the full model, which includes the effect of distance, administrative boundaries and 6-digit IPC match to explore both hypothesis 1 and 2 simultaneously to be able to separate out the effects of institutional and geographical differences. In doing so, while distance and boundary effects remain significant, their strengths appear considerably smaller. Here, a local patent has an increased probability of citing the original patent by 21.7 percentage points, being in the same region as 10.6 percentage points and in the same country as 3.6 percentage points. In terms of distance, moving from 0-25 miles results in a reduction of 8.8 percentage point probability of citation, decreasing by a further 2.0 percentage points from 25-50 to 50-100 miles. After this, there is no more depreciable effect of distance. These results therefore support hypothesis 1 and 2 and suggests that institutional and distance-based effects influence knowledge spillovers simultaneously.

*Table 3*: Baseline Probit Regression margin effects

|  | Administrative boundaries (1) | Distance (2) | Distance boundaries (3) | Baseline (4) | Top quartile patents (5) | Bottom quartile patents (6) |
|---|---|---|---|---|---|---|
| **Boundary match** | | | | | | |
| Local | 0.267 (0.013)*** | | | 0.217 (0.014)*** | 0.184 (0.038)*** | 0.228 (0.026)*** |
| Region | 0.121 (0.009)*** | | | 0.106 (0.010)*** | 0.120 (0.022)*** | 0.123 (0.019)*** |
| Country | 0.054 (0.004)*** | | | 0.036 (0.006)*** | 0.046 (0.012)*** | 0.028 (0.011)** |
| **Distance (miles)** | | | | | | |
| ln(distance+1) | | -0.021 (0.001)*** | | | | |
| 25-50 | | | -0.257 (0.014)*** | -0.088 (0.016)*** | -0.032 (0.044) | -0.113 (0.028)*** |
| 50-100 | | | -0.326 (0.012)*** | -0.108 (0.016)*** | -0.012 (0.042) | -0.111 (0.027)*** |
| 100-250 | | | -0.359 (0.011)*** | -0.094 (0.015)*** | -0.015 (0.040) | -0.102 (0.026)*** |
| 250-500 | | | -0.407 (0.010)*** | -0.117 (0.015)*** | 0.003 (0.040) | -0.144 (0.026)*** |
| 500-1000 | | | -0.410 (0.010)*** | -0.118 (0.015)*** | -0.006 (0.041) | -0.133 (0.026)*** |
| 1000-2500 | | | -0.382 (0.011)*** | -0.098 (0.015)*** | 0.067 (0.041) | -0.170 (0.027)*** |
| 2500-5000 | | | -0.409 (0.010)*** | -0.118 (0.015)*** | 0.035 (0.040) | -0.173 (0.026)*** |
| >5000 | | | -0.415 (0.010)*** | -0.122 (0.015)*** | 0.017 (0.040) | -0.174 (0.026)*** |
| **Technology match** | | | | | | |
| IPC 6 digit match | 0.125 (0.002)*** | 0.129 (0.002)*** | 0.126 (0.002)*** | 0.124 (0.002)*** | 0.075 (0.008)*** | 0.150 (0.004)*** |
| **Observations** | 291044 | 291044 | 291044 | 291044 | 74116 | 74546 |
| **R^2** | 0.0164 | 0.0113 | 0.0148 | 0.0166 | 0.0047 | 0.0312 |
| **log likelihood** | -198430 | -199450 | -198740 | -198390 | -51131 | -50057 |
| **LLR p-value** | (0.000)*** | (0.000)*** | (0.000)*** | (0.000)*** | (0.000)*** | (0.000)*** |

*This table reports the coefficient estimates for the margin effects of a Probit regression model of citations to UK patents. These coefficients represent the expected percentage point increase in the probability of a citation occurring as a result of the dummy variable changing from 0 to 1. This focuses on the results when including administrative boundaries, distance boundaries against a baseline of 0-25 miles, and the quality of patents as measured by the number of citations. Standard errors are reported in parentheses with statistical significant reported at the * 10%, ** 5% and *** 1% level.*

Thus, institutional differences may be separated from distance effects in terms of their influence on knowledge spillovers and this is important going forward. It must also be noted that the effect of a match at the 6 digit IPC level remains positive and significant here, and in all subsequent specifications, supporting the idea that knowledge spills over more readily within the immediate technology class, but also that the effects of distance and institutional similarity are not just reflecting industry similarity and concentration (Belenzon & Schankerman, 2013).

The final two columns build on this by exploring the suggestion that new and innovative pieces of technology are less likely to be affected by distance because of their importance for firms and individuals (Carlino & Kerr, 2015). For this purpose, importance is measured through patent citation count, with higher citations taken to reflect more important knowledge (Almeida & Kogut, 1999). Here, column 5 represents the baseline specification applied to patents in the top 25% of the number of citations received, thus taken as important knowledge. The results show that although border effects are still positive and significant, distance is not important for highly valued knowledge. Furthermore, the strength of the 6-digit IPC match is almost half the baseline result, suggesting that important knowledge is also more likely to spillover to other firms in different industries. Hence, while distance or being in a different industry may not negatively influence the probability of citation, suggesting that important knowledge may spillover widely, being located within the same administrative area and hence sharing institutional backgrounds is likely to positively influence your ability to access that information. Thus, institutional similarities is important for accessing important knowledge. This is in contrast to low valued patents, which show stronger distance decays than the baseline model and that the 6-digit match is also much stronger. Thus, less valued information may be less likely to travel far or to spillover to other industries.

## 5.3 The effects of technology on knowledge spillovers

The previous section however took the entire period of 1977-2019 without considering the effect that technology may have had on these relationships (Charlot & Duranton, 2006). To measure this, the data is initially split into two periods of prior to, and after 1994 in the dataset in columns 1 and 2 of table 5. This is because, to allow patents time to receive citations, the original patent year is limited to 2012 otherwise the results could be biased due to fewer numbers of citations and limited dissemination for any patent originating beyond that time frame (Helmers, 2019). Thus, the median year for originating patents becomes 1994 to which the baseline regression specification is applied to consider how technology may have influenced these relationships.

The first two columns show that from 1994 onwards, knowledge spillovers decay much faster over distance after 1994 than prior to 1994, in contrast to hypothesis 3. This can clearly be seen in the first three distance boundaries of 25-50, 50-100 and 100-250 miles in both column 1 and 2, whereby the depreciable effect on the probability of citation is much stronger for post 1994. However, in contrast the institutional effects as given by the boundary conditions remain of similar strength. This suggests that technology has not affected the influence of institutional boundaries for knowledge spillovers, but has reinforced the effects of distance. In addition to this, the strength of the 6 digit IPC match has decreased by 3.7 percentage points between these periods, suggesting that knowledge is more likely to cross over between industries than it did previously.

However, this relationship can be explored further by breaking down the period from 1977 to 2012 into five different periods in columns 3 to 7. From this, it can be seen that, apart from column 4, the effects of distance and of being in the same local area increase over each period such that the probably of citation at each distance boundary decreases quicker over time, while the effect of being in the same local area increases. This therefore suggests that the importance of distance in

*Table 4*: The effects of distance on citation probability across different time periods from 1977 to 2019

| | Original patents prior to 1994 (1) | Original patents after 1994 (2) | Original patents priority year 1977-84 (3) | Original patents priority year 1984-91 (4) | Original patents priority year 1991-98 (5) | Original patents priority year 1998-2005 (6) | Original patents priority year 2005-12 (7) |
|---|---|---|---|---|---|---|---|
| **Boundary match** | | | | | | | |
| Local | 0.217 (0.019)*** | 0.219 (0.023)*** | 0.137 (0.030)*** | 0.242 (0.026)*** | 0.252 (0.032)*** | 0.188 (0.036)*** | 0.317 (0.048)*** |
| Region | 0.092 (0.014)*** | 0.119 (0.015)*** | 0.096 (0.023)*** | 0.101 (0.019)*** | 0.107 (0.020)*** | 0.118 (0.023)*** | 0.089 (0.033)*** |
| Country | 0.042 (0.008)*** | 0.026 (0.009)*** | 0.052 (0.014)*** | 0.038 (0.011)*** | 0.043 (0.012)*** | 0.010 (0.013) | 0.019 (0.019) |
| **Distance (miles)** | | | | | | | |
| 25-50 | -0.062 (0.021)*** | -0.128 (0.026)*** | -0.114 (0.034)*** | -0.025 (0.030) | -0.117 (0.036)*** | -0.097 (0.041)** | -0.152 (0.055)*** |
| 50-100 | -0.061 (0.020)*** | -0.175 (0.025)*** | -0.094 (0.032)*** | -0.050 (0.029)* | -0.126 (0.035)*** | -0.154 (0.039)*** | -0.204 (0.051)*** |
| 100-250 | -0.066 (0.019)*** | -0.134 (0.024)*** | -0.101 (0.031)*** | -0.042 (0.027) | -0.104 (0.033)*** | -0.141 (0.037)*** | -0.134 (0.048)*** |
| 250-500 | -0.096 (0.020)*** | -0.151 (0.024)*** | -0.131 (0.032)*** | -0.068 (0.028)** | -0.124 (0.034)*** | -0.158 (0.037)*** | -0.160 (0.048)*** |
| 500-1000 | -0.106 (0.020)*** | -0.147 (0.024)*** | -0.146 (0.032)*** | -0.075 (0.028)*** | -0.129 (0.034)*** | -0.155 (0.037)*** | -0.141 (0.048)*** |
| 1000-2500 | -0.088 (0.020)*** | -0.127 (0.024)*** | -0.152 (0.034)*** | -0.053 (0.028)* | -0.112 (0.034)*** | -0.117 (0.037)*** | -0.151 (0.049)*** |
| 2500-5000 | -0.111 (0.020)*** | -0.142 (0.024)*** | -0.162 (0.032)*** | -0.081 (0.028)*** | -0.119 (0.033)*** | -0.154 (0.037)*** | -0.139 (0.048)*** |
| 5000+ | -0.115 (0.020)*** | -0.151 (0.024)*** | -0.151 (0.032)*** | -0.091 (0.028)*** | -0.133 (0.034)*** | -0.161 (0.037)*** | -0.144 (0.048)*** |
| **Technology match** | | | | | | | |
| | 0.143 | 0.106 | 0.171 | 0.138 | 0.107 | 0.109 | 0.097 |

| IPC 6 digit match | (0.003)*** | (0.004)*** | (0.005)*** | (0.004)*** | (0.006)*** | (0.007)*** | (0.010)*** |
|---|---|---|---|---|---|---|---|
| **Observations** | 133676 | 152906 | 41754 | 71080 | 74396 | 62879 | 36460 |
| **R^2** | 0.026 | 0.011 | 0.034 | 0.024 | 0.013 | 0.011 | 0.012 |
| **log likelihood** | -90370 | -104790 | -27955 | -48110 | -50914 | -43124 | -24962 |
| **LLR p-value** | (0.000)*** | (0.000)*** | (0.000)*** | (0.000)*** | (0.000)*** | (0.000)*** | (0.000)*** |

*This table reports the coefficient estimates for the margin effects of a Probit regression model of citations to patents originating in the UK. This focuses on the results using the baseline specification across periods of 1) prior to 1994, 2) after 1994, 3) 1977-1984, 4) 1984-1991, 5) 1991-1998, 6) 1998-2005, 7) 2005-2012. Standard errors are reported in parentheses with statistical significance reported at the \* 10%, \*\* 5% and \*\*\* 1% levels.*

knowledge spillovers has been increasing over time, supporting the conclusion of columns 1 and 2, and in contrast to hypothesis 3.

These results are contrary to findings such as Keller (2002) and Thompson (2006) who find evidence of decreasing localisation of knowledge as a result of technology. This could be potentially be explained by the way in which new technologies are used, as Charlot and Duranton (2006) suggest that instead of technology allowing for communication across longer distances, it could be used to reinforce existing channels of communication. They also suggest that face-to-face communication is unlikely to be replaced by technology for the purpose of complex information transfer. These results are also in line with findings by Sonn and Storper (2008) who show that in the US, the importance of proximity at the Country, State and Metropolitan statistical area has increased over time, attributing this to the shortening of the technology lifecycle and the implications this has on the requirements for quick access to new information. This would therefore provide a potential explanation for the trend seen here, but further research would be required to separate out these effects. The only evidence in support of hypothesis 3 here is the strength of country effects and the match of the 6-digit IPC decreasing over this period, suggesting that improvements in communication technologies have allowed for knowledge to flow across country and technology boundaries easier than before. However, the overall results suggests that technology has reinforced the effects of distance and institutional similarity on knowledge spillovers.

### 5.5 The effects of distance on knowledge spillovers in different industries

Finally, it is important to explore the different effects of that distance, institutions and technology similarity may have across industries, as suggested in hypothesis 4. This is done by performing the baseline regression across different industries given by their IPC classification as defined by the OECD (OECD.stat, 2020)[1]. The results in table 7 below clearly support the hypothesis that knowledge spillovers in different industries. For example, while in biotechnology administrative boundary effects appear strongly positive and significant, distance has no significant effect on how knowledge may be transferred. This therefore suggests that in the biotechnology industry, similarity of local institutions may positively affect your access to new knowledge, but that this is unaffected by distance. This could potentially be due to the importance of knowledge transferability in the medical industry, and the dependence on codification for knowledge transferral, which would reduce the effects of distance, but institutional similarity may allow for quicker adoption of these new technologies if they are embedded in similar systems (Howells, 2002). Furthermore, the match along the 6-digit IPC code is weak, suggesting that knowledge can flow between different areas of technology within biotechnology, potentially because of its linkage with both biology and IT. Similar results can also be seen for the

---
[1] For industry codes used for each industry see Appendix A

medical industry where only local and regional boundaries are seen to positively constrain the flow of knowledge spillovers, suggesting an emphasis on codified knowledge sharing across greater distances but also local standards of practices.

In contrast in the pharmaceutical and IT industries, both distance and administrative boundaries play a role in influencing knowledge spillovers. In the case of pharmaceuticals, local and regional boundaries positively constrain knowledge spillovers, suggesting a strong influence of institutional proximity. Furthermore, at distances of greater than 100 miles, knowledge spillovers are seen to decay, with similar results seen in the IT industry (although significant decay is seen from the 500-1000 mile boundary), suggesting that face-to-face interaction is important. This could be because of less codification and standardisation within these industries, compared to the biotechnology and medical industries, meaning that tacit knowledge sharing is more important. However, further exploration is required to support this conclusion including a deeper dive into the workings of each industry to industry why this may be the case and identifying exactly which scales knowledge spillovers may break down. There are also differences in the strength of the 6 digit IPC matching effect across industries, suggesting that in some, the need to match across similar technologies is more important such as the medical results in contrast to biotechnology. Overall this suggests that the effects of distance on knowledge spillovers, as proxied by patent citations, differs across industries, supporting the results of previous literature (Adams & Jaffe, 1996; Duranton & Overman, 2005; Belenzon & Schankerman, 2013). To understand why though, we need to dig deeper as this can have important consequences for policy support across industries in relation to location.

*Table 5*: The effects of geography across industries

|  | Biotechn-ology (1) | Medical (2) | Pharmac-euticals (3) | IT (4) |
|---|---|---|---|---|
| **Boundary match** | | | | |
| Local | 0.343 | 0.298 | 0.326 | 0.248 |
|  | (0.092)*** | (0.068)*** | (0.050)*** | (0.045)*** |
| Region | 0.162 | 0.096 | 0.072 | 0.053 |
|  | (0.051)*** | (0.041)** | (0.035)** | (0.028)* |
| Country | 0.060 | 0.034 | 0.026 | 0.029 |
|  | (0.028)** | (0.025) | (0.020) | (0.016)* |
| **Distance (miles)** | | | | |
| 25-50 | -0.024 | -0.045 | -0.060 | -0.059 |
|  | (0.114) | (0.074) | (0.058) | (0.048) |
| 50-100 | -0.109 | -0.019 | -0.088 | -0.106 |
|  | (0.107) | (0.070) | (0.058) | (0.046)** |
| 100-250 | -0.052 | -0.035 | -0.110 | -0.089 |
|  | (0.104) | (0.067) | (0.054)** | (0.044)** |
| 250-500 | -0.029 | -0.035 | -0.171 | -0.123 |
|  | (0.104) | (0.068)* | (0.054)*** | (0.044)*** |
| 500-1000 | -0.030 | -0.044 | -0.138 | -0.119 |
|  | (0.104) | (0.068) | (0.055)** | (0.044)*** |
| 1000-2500 | 0.016 | -0.030 | -0.101 | -0.101 |
|  | (0.105) | (0.069) | (0.055)* | (0.045)** |
| 2500-5000 | 0.016 | -0.044 | -0.109 | -0.127 |
|  | (0.104) | (0.068) | (0.054)** | (0.044)*** |
| 5000+ | -0.017 | -0.041 | -0.102 | -0.115 |
|  | (0.104) | (0.068) | (0.054)*** | (0.044)*** |
| **Technology match** | | | | |
| IPC 6 match | 0.063 | 0.158 | 0.085 | 0.086 |
|  | (0.014)*** | (0.010)*** | (0.008)*** | (0.006)*** |
| **Observations** | 11240 | 18044 | 18658 | 41930 |
| **R^2** | 0.010 | 0.018 | 0.019 | 0.009 |
| **log likelihood** | -7711.7 | -12286 | -12683 | -28790 |
| **LLR p-value** | (0.000)*** | (0.000)*** | (0.000)*** | (0.000)*** |

This table reports the coefficient estimates for the margin effects of a Probit Regression model of citations to patents originating in the UK across different industries. Standard errors are reported in parentheses with statistical significance reported at the * 10%, ** 5% and *** 1% level. For patent industry classification see Appendix A.

# 6 Conclusions

This paper explored the role of technological overlap, administrative boundaries and geographical distance on knowledge spillovers for the case of the UK. It showed that, despite technological advances, geography does play an important role, as often reiterated in the economic geography literature. This confirms that the accessibility of a firm to external resources and knowledge predisposes its ability to innovate. Although there are many caveats surrounding patent citation as a proxy for knowledge spillovers, we demonstrated in this work that it can nevertheless provide a reliable method for tracing them at a highly granular geographic level. In addition, it allowed us to separate out the effects of physical distance, technological similarity and institutional influences on knowledge spillovers, extending earlier research in this area, which examined the effects of distance using administrative boundaries as proxies for distance.

What these results show is that distance and administrative boundaries are significant determinants of knowledge spillovers in the UK to the extent that most knowledge spillovers are exhausted in a range of 100 miles, and that the effects of local, regional and country level boundaries, in terms of their institutional influence, remain significant even when distance is accounted for. This suggests that knowledge spillovers are exhausted within an extended commuting boundary and that face-to-face interaction may be important to facilitate the spillover of important knowledge. Supporting interactions beyond this distance could be achieved through mechanisms such as temporary agglomerations, knowledge pipelines or policies that encourage the agglomeration of firms, such as seen in the Biotechnology cluster in Cambridge (Deloitte, 2015).

We also found that despite improvements in communication technologies, the negative effects of distance on the probability of knowledge spilling over has become stronger since 1977, with knowledge spillovers decaying faster with distance in more recent years. This suggests that instead of communication technologies allowing knowledge to be dispersed more widely, they could be reinforcing existing spatial disparities in knowledge transfer by strengthening existing communication networks rather than supporting the development of new ones. This may explain why the agglomeration of firms remains prevalent and suggests that technology may not significantly alter this phenomena in the future. Thus industrial and economic policy must acknowledge the limitations of technology in fostering knowledge spillovers and take into account distance between firms in policies that may influence industrial location.

Finally, we showed that these effects vary by industry, with some industries exhibiting strong localisation of knowledge spillovers relative to others. This could potentially be related to factors such as differing technological lifecycles, difference in patenting behaviour between industries and also reliance on codification for knowledge sharing within these industries (Jaffe, et al., 2000; While, et al., 2004). This means that policy solutions must pay attention to these differences and should adjust according to the industry they are targeting. However, further research is required to be able to understand the factors that may be influencing these relationships, including the level of codification, interaction and the degree of institutional similarity between firms at a distance.

Future research in this area could explore these relationships in more detail for the wider European context. Within the work presented here, there was no geographical restriction placed on citations or control patents, which while the overall distribution of knowledge production was controlled for, results from US based studies suggest that restricting patents to within Europe may influence the relationships discovered. This could also extend into an exploration of the determinants of knowledge spillovers within the wider European context, which could seek to inform the EU's policy of continued economic integration (Barca, 2009). Furthermore, institutional similarity here is explored using the proxy of administrative boundaries at the local, regional and country level, which previous research suggests institutional influences can extend beyond or between these boundaries (McCann & Van Oort, 2019). Thus, future research could attempt to map these institutional influences and examine the scale at which they operate to be able to integrate into these regressions. This could be supported by including cognitive, organisational and social factors into the model to account for other influences on knowledge spillovers that have been recognised in the literature (Boschma, 2005). Finally, the results are interpreted through existing theories and while correlation is explored, causation is not provided. This can come from more developed studies that include the examining of the mechanisms and channels through which distance and institutional factors influence knowledge spillovers directly (Singh & Marx, 2013), such as how knowledge is transferred through different methods of communication.

## Disclosure statement

No potential conflicts of interest was reported by the Authors

# Appendix

Code can be found at: https://github.com/PhilipDW183/Knowledge-spillovers

## Appendix A

These are the IPC codes used to link a patent to a certain industry given by the OECD's specification (OECD.stat, 2020).

Biotech: A01H1, A01H4, A01K67, A61K35/[12-79], A61K(38, 39), A61K48, C02F3/34, C07G(11, 13, 15), C07K(4, 14, 16, 17, 19), C12M, C12N, C12P, C12Q, C40B(10, 40/02-08, 50/06), G01N27/327, G01N33/(53,54,55,57,68,74,76,78,88,92), G06F19/[10-18,20-24]

Pharmaceutical: A61K

Medical: A61B, A61C, A61D, A61F, A61G, A61H, A61J, A61L, A61M, A61N, H05G

(Inaba & Squicciarini, 2017)

IT: H03K, H03L, H03M, H04B1/69-1/719, H04J, H04L (excl. H04L12/14), H04M(3-13,19,99), H04Q, H04B1/00-1/68, H04B1/72-1/76, H04B3-17 (excl. H04B1/59, H04B5, H04B7), H04H, H04B7, H04W (excl. H04W4/24, H04W12), G06F12/14, G06F21, G06K19, G09C, G11C8/20, H04K, H04M1/66-665, H04M1/667-675, H04M1/68-70, H04M1/727, H04N7/167-7/171, H04W12, G06Q20, G07F7/08-12, G07G1/12-1/14, H04L12/14, H04W4/24, G08B1/08, G08B3/10, G08B5/22-38, G08B7/06, G08B13/18-13/196, G08B13/22-26, G08B25, G08B26, G08B27, G08C, G08G1/01-065, H04B1/59, H04B5, G06F5, G06F7, G06F9, G06F11, G06F13, G06F15/00, G06F15/16-15/177, G06F15/18, G06F15/76-15/82, G06F3/06-3/08, G06F12 (excl. G06F12/14), G06F12/14, G06K1-7, G06k13, G11B, G11C (excl. G11C8/20), H04N5/78-5/907, G06F17/30, G06F17/40, G06F17/00, G06F17/10-17/18, G06F17/50, G06F19, G06Q10, G06Q30, G06Q40, G06Q50, G06Q90, G06Q99, G08G (excl. G08G1/01-065, G08G1/0962-0969), G06F17/20-28, G06K9, G06T7, G10L13/027, G10L15, G10L17, G10L25/63, 66, H04M1 (excl. H04M1/66-665, H04M1/667-675, H04M1/68-70, H04M1/727), G06F3/01-3/0489, G06F3/14-3/153, G06F3/16, G06K11, G06T11/80, G08G1/0962-0969, G09B5, G09B7, G09B9, H04N (excl. H04N5/78-5/907, H04N7/167-7/171), G06T1-9 (excl. G06T7), G06T11 (excl. G06T11/80), G06T13, G06T15, G06T17-19, G09G, H04R, H04S, G10L (excl. G10L13/027, G10L15, G10L17, G10L25/63, 66), H03B, H03C, H03D, H03F, H03G, H03H, H03J, H01B11, H01L29-33, H01L21, 25, 27, 34-51, B81B7/02, B82Y10, H01(P, Q), G01S, G01V3, G01V8, G01V15, G06F(00, 05, 09, 12, 13, 18), G06E, G06F1, G06F15/(02, 04, 08-14), G06G7, G06J, G06K(15,17), G06N, H04M15, H04M17